\documentclass[useAMS,usenatbib]{mn2e}
\usepackage[dvips]{graphicx}

\title[Methanol masers in a candidate circumstellar disc]
      {European VLBI Network observations of 6.7-GHz methanol masers in a candidate circumstellar disc}
\author[L.~Harvey-Smith, R.~Soria-Ruiz]{L.~Harvey-Smith$^{1}$\thanks{E-mail:
      lhs@usyd.edu.au}, R.~Soria-Ruiz$^{2}$\\
$^{1}$Institute of Astronomy, School of Physics, University of Sydney, 2006 NSW, Australia.\\
$^{2}$Observatorio Astron\'{o}mico Nacional, c/ Alfonso XII 3, 28014 Madrid, Spain.}

\begin{document}

\date{\today}

\pagerange{\pageref{firstpage}--\pageref{lastpage}} \pubyear{2008}

\maketitle

\label{firstpage}

\begin{abstract}
The first high-resolution ($\sim$5~mas) VLBI observations of 6.7-GHz methanol masers in DR21(OH)N, a candidate circumstellar disc around a very young massive star, are presented. Previous observations of these masers at 50~mas angular resolution revealed a rotating structure at the position of a candidate massive protostar, with a well-sampled position-velocity diagram suggesting Keplerian rotation. Observations presented here using the European VLBI Network (EVN) have provided the first high angular resolution maps of the masers, providing a test for the disc hypothesis and the Gaussian centroiding technique. The EVN maps have confirmed the shape of the disc and its rotation curve. Weaker maser emission seen previously with MERLIN between the two main spectral peaks is seen in the EVN total power spectrum, but is absent in the cross-power spectrum. This suggests that the spatially extended emission is resolved out by the EVN. The rotating disc is coincident with a Class I massive (proto)star and at the implied centre of an outflow traced by two bow shocks. We discuss the impact of this result on the massive stellar accretion disc hypothesis and on the validity of the centroiding technique to determine the structures of unresolved masers using compact radio interferometric arrays.      

\end{abstract}


\section{Introduction} 

Massive stellar accretion is an important astrophysical process, the mechanics of which are still poorly understood (for a thorough review of the topic, see Zinnecker \& Yorke, 2007). Observing masers in massive star-forming regions allows us to trace the bulk motions of molecular gas in the vicinity of nascent massive stars. This can give us important information about the infall, outflow and rotational motions associated with massive star formation, in particular identifying possible circumstellar discs around massive protostars (Cesaroni et al. 2007).  Recent observations of 6.7-GHz methanol masers using the Multi-Element Radio Linked Interferometer Network (MERLIN) in the massive star-forming region DR21(OH)N led to the discovery of a highly unusual maser with a double peaked spectrum (Harvey-Smith et al. 2008). The maser region is elongated in an east-west direction and has an estimated diameter of 60~au. The resulting position-velocity diagram showed a clear Keplerian rotation profile, suggesting that the methanol masers may be tracing a rotating disc of material falling onto a massive star. The masers coincide with the very young Class I candidate massive star in the infra-red and dust core ERO3 (Marston et al. 2004). This extremely reddened object is at the centre of (and perpendicular to) a possible molecular outflow described by Davis et al. (2007). The magnetic field direction implied from the linear polarization of the methanol masers is roughly perpendicular to the implied plane of the disc. 
The indirect evidence for a young, massive circumstellar disc in DR21(OH)N is considerable, but to remove any uncertainty we must be able to confirm the rotation curve observed by Harvey-Smith et al. (2008). This is the single piece of observational evidence that can (i) directly demonstrate the motion of molecular gas around the young massive star and (ii) produce an estimate of the current mass of the star. In order to test the reliability of the position-velocity diagram and therefore the disc interpretation, the 6.7-GHz methanol masers in DR21(OH)N were re-observed using the European VLBI Network at the same spectral resolution but with an angular resolution ten times greater than that of MERLIN (5~mas compared with 50~mas). In doing this, it would become possible to detect the effect of spatial blending in the MERLIN centroid maps and to verify (or otherwise) the velocity curve of the maser disc.

\section{Observations}

The star forming region DR21(OH)N was observed at 6.7-GHz using the European VLBI Network antennas at Cambridge, Jodrell Bank, Effelsberg, Medicina, Onsala, Torun, Noto and Westerbork. The total observing bandwidth of 0.5 MHz was split into 512 spectral channels, giving a velocity resolution of 0.04 km~s$^{-1}$. The calibrator source J2007+4029 was used as a fringe finder and observations of J2202+4216 were used to calibrate the bandpass. A total of 238 minutes were spent on the target source, leading to an RMS noise level of 60 mJy~beam$^{-1}$. 

Initial flagging and amplitude calibration was carried out by the EVN pipeline (Reynolds et al. 2002). Further manual flagging using the {\sc AIPS} tasks {\sc IBLED} and {\sc SPFLG} was then used to remove bad data. Data from the Cambridge antenna were incorrectly recorded and had to be deleted completely. Delay residuals and their time derivatives were calibrated using {\sc FRING} and bandpass corrections applied with {\sc BPASS}. Using {\sc IMAGR} the dirty map was produced and then self-calibration was applied to the image, using a bright channel with simple point-like maser emission as a model. 

An image cube was made, with each image plane (corresponding to a single spectral channel) being deconvolved using the Hogb\"{o}m {\sc CLEAN} algorithm. Finally, the calibrated image cubes were analysed using the interactive {\sc AIPS} routine {\sc ORFIT}, which fits 2-dimensional Gaussian components to each spectral channel and finds the position of the emission centroids. A lower limit cutoff of 5$\sigma_{RMS}$ was chosen, to avoid the inclusion of spurious features. In the following section, the results of very high-resolution EVN observations of DR21(OH)N are presented.

\section{Results}

The images of DR21(OH)N made with the EVN show methanol masers at the same position, and with a very similar position-velocity curve, to the results of Harvey-Smith et al. (2008). This close agreement confirms the existence of a rotating disc of molecular gas at the postition of the ERO3 protostar. Figure 1 shows the EVN spectrum of 6.7-GHz methanol maser emission in DR21(OH)N. The flux drops to the RMS noise level between the two spectral peaks, suggesting that the extended `pedestal' emission seen by Harvey-Smith et al. (2008) is resolved out with the 5~mas resolution of the EVN.

The positions of 6.7-GHz methanol maser emission centroids in each spectral channel are shown in Figure 2 (top). A signal-to-noise ratio cutoff of 5$\sigma_{RMS}$ was used. Figure 2 (bottom) shows a comparison between the EVN centroids and the MERLIN centroids from Harvey-Smith et al. (2008). There is a clear correspondence between the shapes of the emission around the two bright peaks, although the EVN maps show the spectral peaks more spatially separated than the MERLIN data. This is probably due to the MERLIN observations being spatially unresolved and therefore the centroids of one maser peak being slightly offset towards the position of the other `competing' maser (competitive smearing).  The obvious difference between the results is that the EVN observations show far fewer spectral channels with emission above  5$\sigma_{RMS}$, that is to say that the weaker flux \emph{between} the two maser peaks is not seen by the EVN. The most likely cause of this is MERLIN's superior surface brightness sensitivity to low brightness temperature (spatially extended) emission, due to its shorter baselines. Another possible explanation is that the RMS noise level of the EVN images is significantly higher than the MERLIN images. These issues will be discussed in more detail in section 4.

The position-velocity diagram of the maser feature is shown in Figure 3 (top), where the impact parameter is plotted in the east-west direction across the disc. This is compared with the MERLIN results from Harvey-Smith et al. (2008) in Figure 3 (bottom). The EVN results show qualitatively a very similar rotation curve to the previously published MERLIN data, albeit with a greater implied disk radius from the EVN observations. The reasons for the differences between the two curves are vital in interpreting the nature of the region. The implications of these results will be discussed in the following section.

\begin{figure}
\includegraphics[angle=-90,scale=.35]{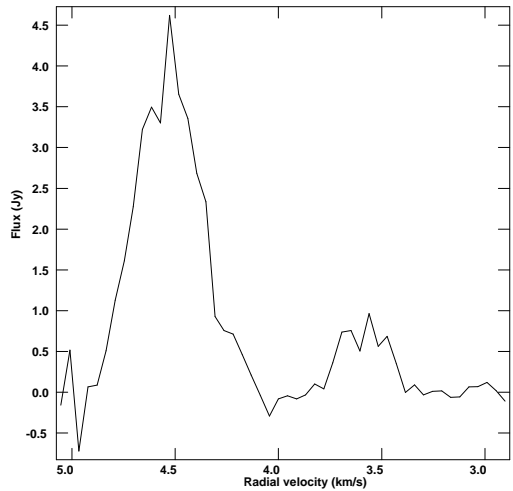}
\caption{The EVN cross-power spectrum of DR21(OH)N, showing the two maser emission peaks at 3.5 and 4.5 km~s$^{-1}$.}
\end{figure}

\begin{figure*}
\includegraphics[angle=-90,scale=.5]{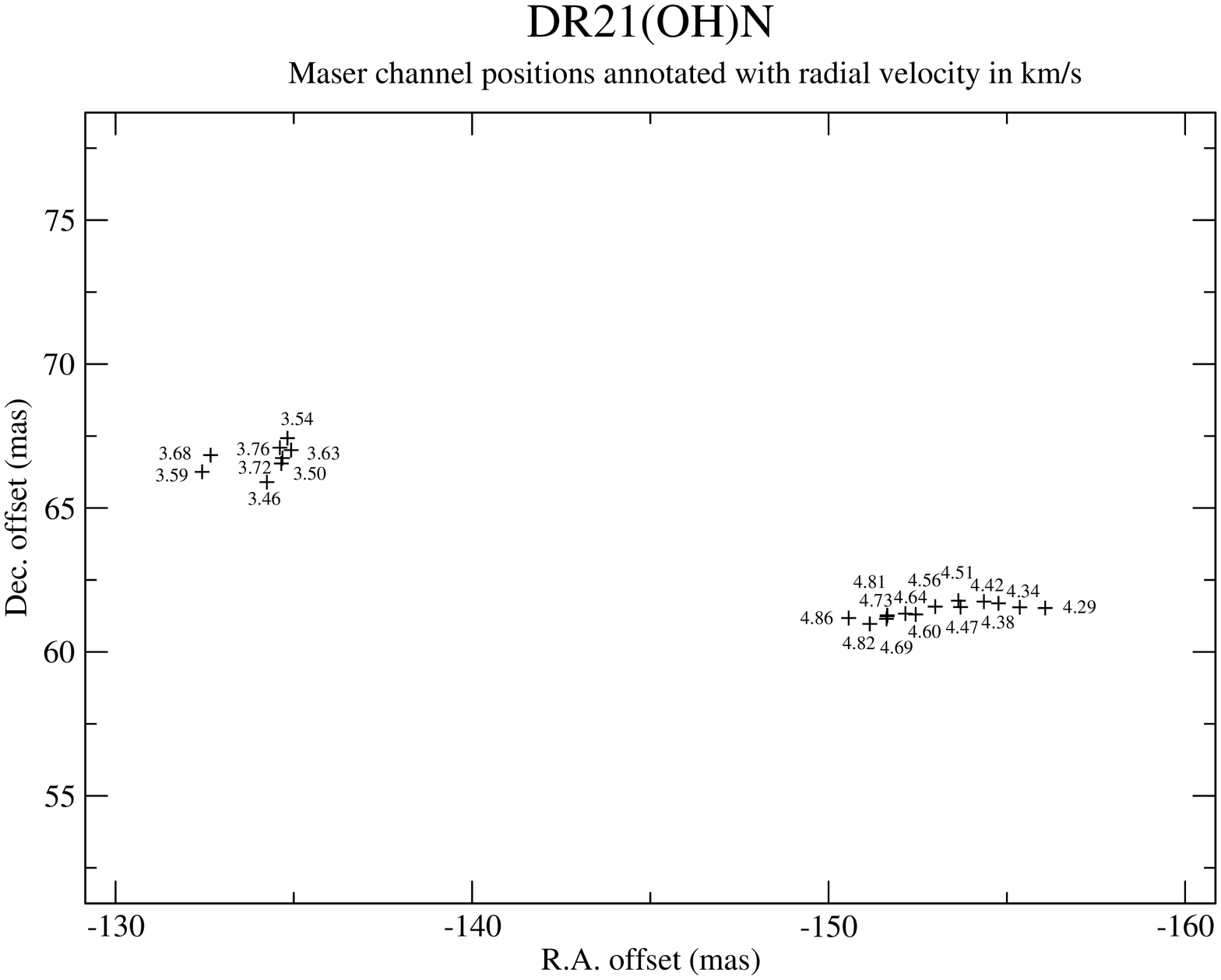}
\includegraphics[angle=-90,scale=.5]{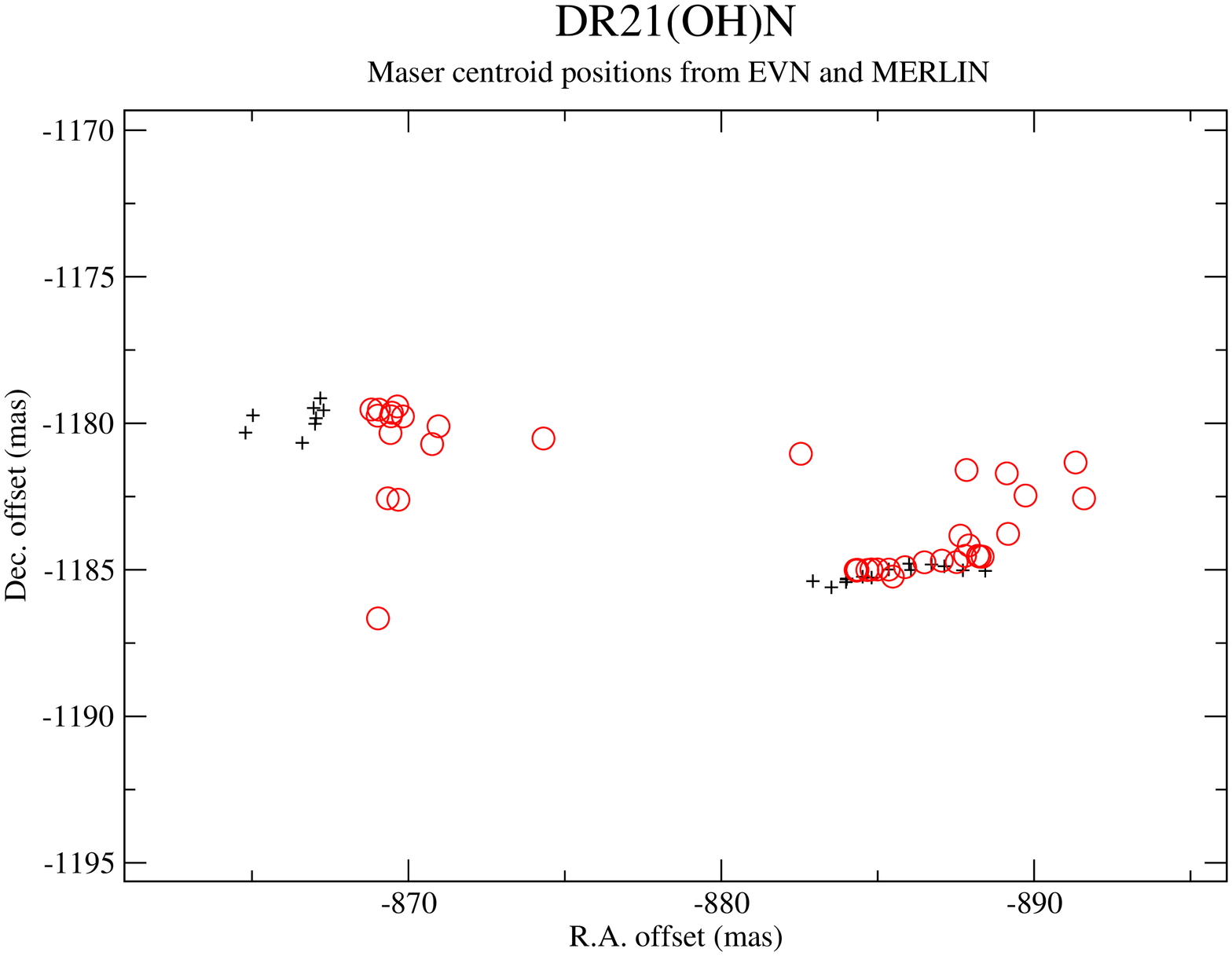}
\caption{Top: Positions of 6.7-GHz methanol maser emission centroids (crosses) in DR21(OH)N observed with the EVN. The symbols are annotated with the central velocity of the channel in kilometers per second. Bottom: Comparison of the positions of 6.7-GHz methanol maser centroids, observed with the EVN (crosses: this paper) and MERLIN (circles: Harvey-Smith et al. 2008). The maps were registered by aligning the brightest spectral channel in each data set. The reference position is RA: 20$^h$39$^m00.457^s$, Dec: 42\degr24\arcmin38\arcsec.330. }
\end{figure*}

Harvey-Smith et al. (2008) made maps of DR21(OH)N with MERLIN, using Gaussian centroiding of each spectral channel. The use of this technique gave the observations an effective angular resolution far greater than that afforded by the 50~mas primary beam of MERLIN, but also made the maps susceptible to confusion effects. As pointed out by Weintroub et al. (2008), the centroiding technique has some possible pitfalls when used to map a source with multiple, spatially unresolved components. This may result in a map with points tracing a blended combination of several independent structures, which would mean that the centroid positions of the masers were misleading. As the physical interpretation of the methanol masers in DR21(OH)N rests largely on the shape of the position-velocity diagram, this type of ambiguity caused by centroiding blended emission is clearly something that has to be eliminated. In the following sections, the implications of the new EVN results on the likely origins of the 6.7-GHz methanol masers in DR21(OH)N are discussed.   

\section{Discussion}

\subsection{Is the maser emission over-resolved by the EVN?}

 With an angular resolution of 5~mas at 6.7 GHz, the EVN is able to probe the internal velocity structure of the methanol masers in DR21(OH)N at very high angular resolution. This allowed us to determine the position of the emission centroids in each of 22 adjacent velocity channels. Given the resolved nature of the two spectral peaks in position and velocity, each centroid should be considered independent as no deviation of the centroid position would be expected from the neighbouring peak.

\begin{figure*}
\includegraphics[angle=-90,scale=.5]{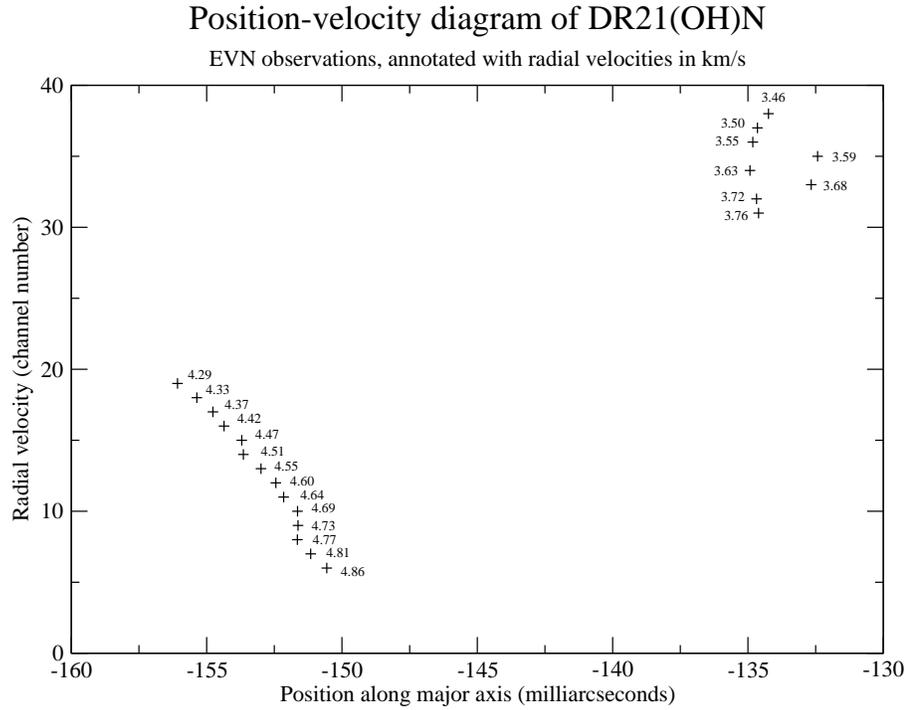}
\includegraphics[angle=-90,scale=.5]{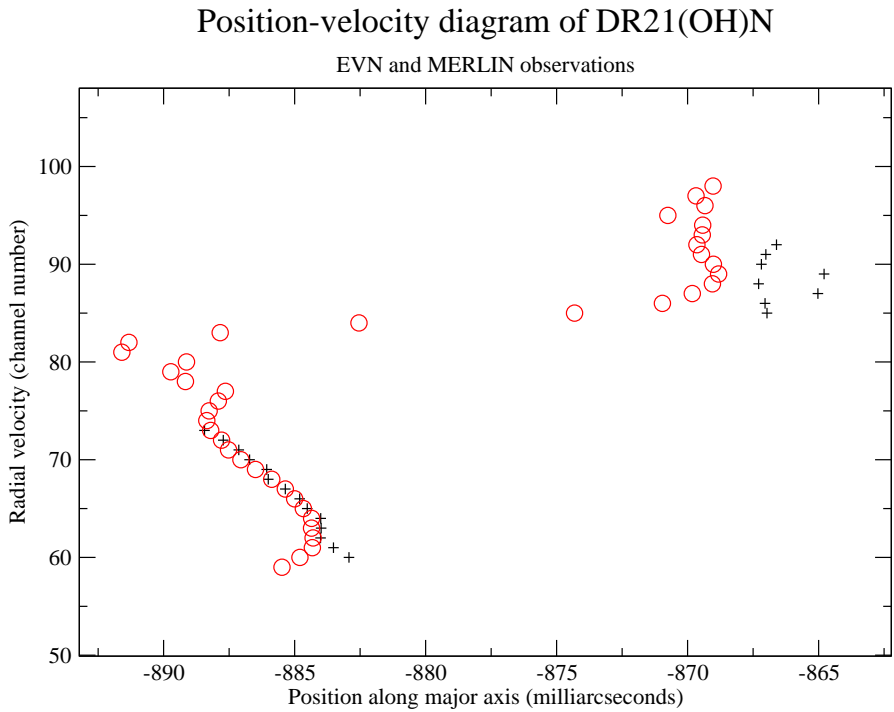}
\caption{Top: Position-velocity diagram for the 6.7-GHz methanol maser emission in DR21(OH)N (crosses) with position being in an east-west direction. The EVN centroids are annotated with the central velocity of each channel in kilometers per second. Bottom: A comparison of the position-velocity plots of the EVN (crosses: this paper) and MERLIN (circles: Harvey-Smith et al. 2008) data in DR21(OH)N, showing the general agreement between the two velocity curves. }
\end{figure*}

The internal shape and velocity structure of masers is of fundamental importance as it enables us to understand the morphology and kinematics of the molecular cloud that is producing the maser. This is not only interesting in terms of the physical origin of the masers, but also as a probe of the overall gas motions in the region. For this reason, it is vital to understand the complexities resulting from the data analysis techniques employed and to compare results using different observing methods. 

By comparing the results from Harvey-Smith et al. (2008) and EVN observations with ten times greater angular resolution, it has been possible to verify the internal velocity structure of the brightest 6.7-GHz methanol maser emission in DR21(OH)N, corresponding to the approximately $v \propto 1/r^{1/2}$ (Keplerian) portions of the curve (Figure 2). The most obvious difference between the two results is that the EVN observations recorded a 5$\sigma_{RMS}$ maser detection in only 22 channels as opposed to 39 channels in the MERLIN observations (with both observations having the same velocity resolution). The weaker emission between the two peaks, presumably associated with the near-side outer edge of the disc, was not detected with the EVN. This could be due to one of three reasons. 

The first is that the EVN images were too noisy to show the emission from the outer ring. In response to this suggestion, the RMS noise levels in the EVN maps are typically 8 times the noise level in the MERLIN maps. This lends support to the notion that the weaker emission from between the main peaks could not be seen above the background noise.

The second possibility is that the centroiding technique applied to the MERLIN data gave average positions of spatially blended components, rather than actual positions of maser emission. The channels where maser emission is apparently linking the two maser peaks may simply be channels where the 50~mas synthesised beam of MERLIN is giving rise to a single centroid position that lies between the two true maser spots. This can be tested by examining the exact shape of the position-velocity curve of the methanol maser centroids determined with the MERLIN data. The shape of the MERLIN position-velocity diagram in Figure 3 (bottom) would be very difficult to explain in terms of the effects of centroiding two spatially-separated but unresolved masers. In each velocity channel the weaker emission in the wings of one spectral line would be expected to move directly towards the peak centroid of the other spectral line. This is not what is seen in figure 3. In fact, close to the linear portion of the position-velocity curve the centroids move back across to a maximum separation along the major axis before moving towards the `central' position along the east-west major axis. This is what is expected from Keplerian rotation through an edge-on disc. This occurs at both the eastern and western extremes of the region and cannot easily be explained in terms of systematic errors caused by centroiding two unresolved components.

The third possibility is that this fainter radiation from between the spectral peaks is diffuse maser emission with a larger angular size than the brighter emission from the spectral peaks, analogous to the diffuse methanol masers found in W3(OH) (Harvey-Smith \& Cohen 2006). The maser emission from the near-side edge of a circumstellar disc could be the result of weakly-amplified background emission from throughout the disc. In this case, the outer edge disc masers would be visible with MERLIN but not the EVN, as MERLIN is more sensitive to spatially extended emission. The evidence for this suggestion is very strong indeed. An unambiguous way to determine whether there is spatially unresolved emission in an interferometric observation is to compare the total power spectrum with the cross-correlation spectrum. Figure 4 allows a comparison of the total power spectrum from the Effelsberg telescope - which shows a significant portion of unresolved flux between the the peaks - with the cross-correlation spectrum. This plot demonstrates that the cross-correlated flux returns to the noise level between the peaks, which clearly demonstrates that there is significant flux between the two maser spots that is resolved out by the EVN. 

\begin{figure}
\includegraphics[angle=-90,scale=.3]{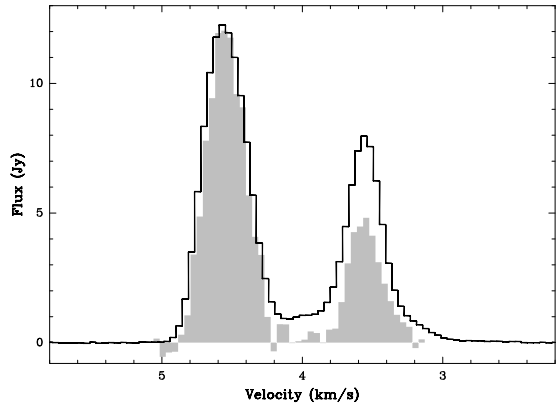}
\caption{Total power spectrum from the Effelsberg telescope (solid line) versus the cross-power spectrum (greyscale) in DR21(OH)N. This comparison shows the extended emission between the two spectral peaks that was resolved out by the EVN. The reference position is RA: 20$^h$39$^m00.457^s$, Dec: 42\degr24\arcmin38\arcsec.330.}
\end{figure}

\subsection{ERO3: A Class I massive protostar with a possible outflow}

The methanol 6.7-GHz masers that trace a possible disc in DR21(OH)N coincide with ERO3, which was identified as a likely candidate for a young massive star by Marston et al. (2004). They observed the DR21 region using the Infra-Red Array Camera (IRAC) and the Multiband Imaging Photometer for Spitzer (MIPS) on the Spitzer Space Telescope. They identified several EROs and calculated their spectral indices based on Spitzer and 2MASS data. On the basis of the spectral index, they classified ERO3 as a Class I protostar. This class of protostars are still embedded in their circumstellar envelopes, but have central clump masses that dominate over the circumstellar envelope mass (Lada 1987).

An intriguing observation has been made of two chains of bright H$_2$ knots, approximately equidistant, to the north and south of ERO3 (Davis et al. 2007; Appendix, Figure A2 jets 6 and 9). One scenario put forward by Davis and co-workers was that these features are bow shocks caused by a north-south outflow originating from ERO3. This fits very well with the methanol maser observations presented here, as the outflow is perpendicular to the proposed circumstellar disc and is also aligned approximately parallel to the magnetic field direction implied from the polarization of the methanol masers in the disc (Harvey-Smith et al. 2008). The precise origin of the outflow cannot be verified without direct observations of the region at high angular resolution.
 
Several authors have derived the physical conditions within the DR21(OH)N region from studying the molecular gas and dust. Jakob et al. (2007) carried out sub-mm and Far Infra-Red (FIR) observations in DR21(OH)N. The source they designated DR21 FIR1 coincides with a 1.3-mm dust core (Chandler et al. 1993). Jakob and co-workers quoted a dust temperature of 33~K for the region, which was derived using a dust continuum fit over the whole source, with much poorer angular resolution than either MERLIN or the EVN. They placed an upper limit of 82~K on the gas temperature because the high-$J$ CO was not detected in DR21 FIR1. Chandler et al. (1993) derived a dust temperature of 65~K in DR21 FIR1 from a fit to the dust continuum spectrum.  We can compare these values with the theoretical model of Cragg et al. (2005), who calculated the brightness temperature of several Class II methanol masers as a function of the temperature, density and abundance of gas and dust in the region. 

According to the Cragg et al. (2005) model, which includes pumping by far infra-red radiation and by collisions, the brightness temperatures of class II methanol masers depend very little on the temperature of the gas in the region. The influence of the dust temperature is very strong, with the models showing an initiation of 6.7-GHz methanol masers at T$_{dust}$=110~K, below which no masers are produced. The model allows masers to be produced in regions with gas densities as determined by Jakob et al. (2007), but is inconsistent with the dust temperature estimates for DR21 FIR1 of 33~K and 65~K. As these dust temperatures are beam-averaged and the ISO Long Wavelength Spectrometer beam is much larger than the maser region ($\sim$7 arcminutes compared with a few milliarcseconds), the temperature in the maser region is probably much higher than those calculated by Jakob et al. (2007) and Chandler et al. (1993).

\subsection{Are the methanol masers tracing a massive circumstellar disc?} \label{disc}

The appearance and velocity gradient of the methanol maser feature in DR21(OH)N is quite different to most masers in this and other massive star-forming regions. The maser feature in DR21(OH)N has an extraordinarily large and ordered velocity gradient of 32 m~s$^{-1}~au^{-1}$. This is much larger than the unusual broadline region of W3(OH) (Harvey-Smith \& Cohen 2006) and the circumstellar disc in NGC 7538 (Pestalozzi et al. 2004 and Erratum), both of which have velocity gradients of $~$12 m~s$^{-1}~au^{-1}$. This, coupled with the striking resemblance of the rotation curve to a Keplerian profile seems like sufficient evidence to suggest that the methanol maser feature is a circumstellar disc.  

Are there alternative explanations for the origin of the methanol masers in DR21(OH)N? If, as widely thought, methanol 6.7-GHz masers are \emph{always} associated with massive star-formation then it is difficult to find an explanation for this rotating structure other than a clump of gas orbiting a massive star. One possibility is that the masers originate from gas orbiting a lower mass star. This is unlikely because 6.7-GHz methanol masers are believed to require a minimum far infra-red radiation flux as an ongoing pumping mechanism to sustain the population inversion, which requires close proximity to a source of energy such as an O or B-type star. Due to a lack of any observational evidence to the contrary it seems that the energy output of lower-mass stars is not sufficient to produce methanol 6.7-GHz masers. 

In order to answer the question of whether this is truly a massive circumstellar disc, we must first estimate the mass of the star by modeling the maser emission and the inclination angle of the disc. Such a model is currently in development by Harvey-Smith et al. (\emph{in prep.}).

\section{Conclusions}

Observations of the DR21(OH)N region using the European VLBI Network have confirmed the existence of an unusual methanol maser feature at 6.7~GHz with a double-peaked spectral profile and a velocity gradient of 32 m~s$^{-1}~au^{-1}$. Analysis of the single dish spectrum reveals weaker maser emission linking the two peaks. Gaussian centroid analysis of the maser emission in each spectral channel revealed a velocity profile resembling a Keplerian rotation curve. The masers coincide with a Class I (very young) B-type protostar, which is located at the centre of an outflow seen by the bow-shocks in H$_2$. The ouflow lies parallel to the magnetic field direction in the maser disc, implied by observations of the linear polarization of the methanol masers. If this object is not a disc around a massive protostar then it would be very difficult to explain the origin of the large velocity gradient and clear Keplerian rotation profile. A model is currently under development to constrain the central mass of the protostar in DR21(OH)N, which will prove very important in verifying the origin of this disc, resulting in wider implications in testing current models of massive star-formation .

\section*{Acknowledgments}
LHS thanks Simon Ellingsen, John Dickey and Jim Moran for useful discussions regarding the centroiding technique and Gemma Anderson for general comments on the manuscript. The authors thank staff at the Joint Institute for VLBI in Europe for their excellent support of the EVN observations and the anonymous referee for their helpful comments. The European VLBI Network is a joint facility of European, Chinese, South African and other radio astronomy institutes funded by their national research councils

\label{lastpage}

\end{document}